\documentclass{amsart}
\usepackage{amsmath}
\usepackage{amssymb}
\input epsf
\newcommand{\si}[1]{\sigma_{#1}}
\newcommand{\sa}[2]{\sigma_{#1}^{#2}}

\newcommand{\ro}{\rho}
\newcommand{\la}{\lambda}

\newcommand{\g}{\gamma_{0}}
\newcommand{\ga}{\gamma}

\newcommand{\I}{\mathbb I}
\newcommand{\ket}[1]{|{#1}\rangle}
\newcommand{\bra}[1]{\langle {#1} |}
\newcommand{\cH}{{\mathcal H}}
\newcommand{\C}{\mathbb C}
\newcommand{\R}{\mathbb R}
\newcommand{\fA}{\mathfrak A}
\newcommand{\fE}{\mathcal E}
\newcommand{\tr}{\mathrm{tr}\,}
\newcommand{\ptr}[1]{\mathrm{tr}_{#1}}
\newcommand{\tl}[1]{\boldsymbol #1}
\newcommand{\mr}[1]{\mathrm{#1}}

\begin{document}
\begin{center}
\begin{LARGE}
\textbf{Entanglement and nonlocality versus spontaneous emission\\[6mm]
in two -- atom system}
\end{LARGE}\\[12mm]
L. Jak{\'o}bczyk and A. Jamr{\'o}z\\[2mm]
Institute of Theoretical Physics\\ University of
Wroc{\l}aw\\
Pl. M. Borna 9, 50-204 Wroc{\l}aw, Poland
\end{center}
\vskip 12mm
\noindent
{\sc Abstract}: We study evolution of entanglement of two
two-level atoms in the presence of dissipation caused by
spontaneous emission. We find explicit formulas for the amount of
entanglement as a function of time, in the case of destruction of the initial entanglement
and possible creation of a transient entanglement between atoms. We also discuss
how spontaneous emission influences nonlocality of states
expressed by violation of Bell - CHSH inequality. It is shown that
evolving system very quickly becomes local, even if entanglement
is still present or produced.
\section{Introduction}
The process of spontaneous emission by a system of two - level
atoms was extensively studied by several authors
(see e.g. \cite{Dicke,dill,lemb1,lemb2}).  In particular, in the case of
spontaneous emission by two trapped atoms separated by a distance
small compared to the radiation wavelength,  where is a
substantial probability that a photon emitted by one atom will be
absorbed by the other, there are states of the system in which
photon exchange can enhance or diminish spontaneous decay rates.
The states with enhanced decay rate are called superradiant and
analogously states with diminished decay rate are called
subradiant \cite{Dicke}. It was also shown by Dicke, that the
system of two coupled two-level atoms can be treated as a single
four-level system with modified decay rates.
\par
Another aspects of the model of the spontaneous emission
are studied in the present paper.
When the compound system of two atoms is in an entangled state,
the irreversible process of radiative decay usually destroys
correlations and the state becomes unentangled. In the model
studied here, the photon exchange  produces correlations between
atoms which can partially overcome  decoherence caused by
spontaneous radiation. As a result, some amount of entanglement
can survive, and moreover there is a possibility that this
process can entangle separable states of two atoms. The idea that
dissipation can create entanglement in physical
systems, was recently developed in several papers
\cite{plenio,kim,milb,kni}. In particular, the effect of
spontaneous emission on destruction and production of entanglement
in  two - atom system was discussed \cite{J,FT3,FT2}. Possible
production of robust entanglement for closely separated atoms was shown
in Ref. \cite{J}, and the existence of transient entanglement
induced by this process in a system of
two atoms separated by an arbitrary distance was also studied
\cite{FT3,FT2}.
\par
In this paper we also concentrate on arbitrarily separated atoms
and consider the dynamics of entanglement. Similarly as in
\cite{FT2} we take some class of initial states, including
interesting pure and mixed states, and discuss in details its time
evolution as well as the evolution of its entanglement. Note that
our initial states are different from that considered in Ref.
\cite{FT2}. We study also the interesting problem how dissipative
process of spontaneous emission influences nonlocal properties of
initial states. Nonlocality of quantum theory manifets by
violation of  Bell inequalities, and in the case of two two-level
systems it can be quantified by some numerical parameter ranging
from $0$  for local states to $1$ for states maximally violating
some Bell inequality. Atomic dynamics studied in the paper
enables also to consider time evolution of this parameter.
\par
The model considered in the present paper consist of two two-level
atoms coupled to a common thermostat at zero temperature and the
reduced dynamics (in the Markovian approximation) is given by the
semi-group $\{ T_{t} \}_{t\geq 0}$ of completely positive linear
mappings acting on density matrices (see e.g. \cite{Alicki}). The
dynamics takes into account only spontaneous emission and possible
photon exchange between atoms \cite{Agar,FT1}, and the generator
$L_{D}$ of $\{ T_{t} \}_{t\geq 0}$ is parametrized in terms of the
spontaneous emission rate of the single atom $\g$ and the photon
exchange rate $\ga$. In the case of atoms separated by an
arbitrary distance $R$, $\ga$ is strictly smaller then $\g$ and
one can check that the relaxation process brings all initial
states into the unique asymptotic state when both the atoms are
in their ground states. In contrast to the small separation
regime ($\ga=\g$) studied in Ref. \cite{J}, where the robust
entanglement of non-trivial asymptotic states can be analysed, in
the present case only the transient entanglement of some states
can exist. To consider transient entanglement we need to know in
details time evolution of initial states, not only its asymptotic
behaviour, so the analysis of possible generation of entanglement
is much more involved.
\par
In this  paper we calculate  time evolution of an arbitrary
initial density matrix. To obtain an analytic expression for
entanglement
 as a function of time, we concentrate on the class of states
which is left invariant during the evolution, and admits
explicite formula for the measure of entanglement. Next we discuss
in details how evolve pure initial states, both unentangled and
entangled. We show that entanglement as a function of time can
behave very differently depending on initial conditions: it may
monotonically decrease to zero, increase to maximal value and
then decrease to zero or even it can have local minimum and
maximum. In particular, there are  states for which induced
transient entanglement is larger then initial entanglement. Our
solution enables  also to study nonlocality of the evolving
system of two atoms. It turns out that the natural measure of
nonlocality very quickly becomes equal to zero, even if
entanglement is increasing in some time interval.
\newpage
\noindent
\section{Entanglement and Nonlocality for a Pair of Two-Level
Atoms}
\subsection{Measure of entanglement}
\noindent
Consider two-level atom $A$ with ground state $\ket{0}$
and excited state $\ket{1}$. This quantum system can be described
in terms of the Hilbert space $\cH_{A}=\C^{2}$ and the algebra
$\fA_{A}$ of $2\times 2$ complex matrices. If we identify
$\ket{1}$ and $\ket{0}$ with vectors $\bigl( \begin{smallmatrix}
1\\0
\end{smallmatrix}\bigr)$ and $\bigl( \begin{smallmatrix} 0 \\ 1
\end{smallmatrix}\bigr)$ respectively, then the raising and
lowering operators $\si{+},\;\si{-}$ defined by
\begin{equation}
\si{+}=\ket{1}\bra{0},\quad \si{-}=\ket{0}\bra{1}
\end{equation}
can be
expressed in terms of Pauli matrices $\si{1},\; \si{2}$
\begin{equation}
\si{+}=\frac{1}{2}\,(\si{1}+i\,\si{2}),\quad
\si{-}=\frac{1}{2}\,(\si{1}-i\,\si{2})
\end{equation}
For a joint system $AB$
of two two-level atoms $A$ and $B$, the algebra $\fA_{AB}$ is
equal to $4\times 4$ complex matrices and the Hilbert space
$\cH_{AB}=\cH_{A}\otimes \cH_{B}=\C^{4}$. Let $\fE_{AB}$ be the
set of all states of the compound system i.e.
\begin{equation} \fE_{AB}=\{
\ro\in \fA_{AB}\, : \, \ro\geq 0\quad\text{and}\quad\tr \ro =1 \}
\end{equation}
The state $\ro\in \fE_{AB}$ is \textit{separable}
\cite{Werner}, if it has the form
\begin{equation}
\ro=\sum\limits_{k}\la_{k}\ro_{k}^{A}\otimes \ro_{k}^{B},\quad
\ro_{k}^{A}\in \fE_{A},\;\ro_{k}^{B}\in \fE_{B},\; \la_{k}\geq
0\quad\text{and}\quad \sum\limits_{k}\la_{k}=1
\end{equation}
The set
$\fE_{AB}^{\,\rm sep}$ of all separable states forms a convex
subset of $\fE_{AB}$. When $\ro$ is not separable, it is called
\textit{inseparable} or \textit{entangled}. Thus
\begin{equation}
\fE_{AB}^{\,\rm ent}=\fE_{AB}\setminus \fE_{AB}^{\,\rm sep}
\end{equation}
As a measure of the amount of entanglement a given state contains
we take the entanglement of formation \cite{Bennett}
\begin{equation}
E(\ro)=\min \, \sum\limits_{k}\la_{k}E(P_{k})
\end{equation}
where the minimum is taken over all possible decompositions
\begin{equation}
\ro=\sum\limits_{k}\la_{k}P_{k}
\end{equation}
and
\begin{equation}
E(P)=-\tr[ (\ptr{A}P)\,\log_{2}\, (\ptr{A}P)]
\end{equation}
In the case of two two-level atoms, $E(\ro)$  is the function of another useful
quantity $C(\ro)$ called \textit{concurrence}, which also can be
taken as a measure of entanglement \cite{HW, W}. Now we pass to
the definition of $C(\ro)$. Let
\begin{equation}
\ro^{\dag}=(\si{2}\otimes
\si{2})\,\overline{\ro}\,(\si{2}\otimes \si{2})
\end{equation}
where
$\overline{\ro}$ is the complex conjugation of the matrix $\ro$.
Define also
\begin{equation}
\widehat{\ro}=(\ro^{1/2}\ro^{\dag}\ro^{1/2})^{1/2}
\end{equation}
Then the
concurrence $C(\ro)$ is given by \cite{HW,W}
\begin{equation} C(\ro)=\max\;
(\,0, 2p_{\mathrm{max}}(\widehat{\ro})-\tr \widehat{\ro}\,)
\end{equation}
where $p_{\mathrm{max}}(\widehat{\ro})$ denotes the maximal
eigenvalue of $\widehat{\ro}$. The value of the number $C(\ro)$
varies from $0$ for separable states, to $1$ for maximally
entangled pure states.
\par
Consider now the class of density matrices $\ro$
\begin{equation}
\ro=\begin{pmatrix}
0&0&0&0\\
0&\ro_{22}&\ro_{23}&\ro_{24}\\
0&\ro_{32}&\ro_{33}&\ro_{34}\\
0&\ro_{42}&\ro_{43}&\ro_{44}
\end{pmatrix}
\end{equation}
where the matrix elements are taken with
respect to the basis $\ket{1}\otimes\ket{1},\,
\ket{1}\otimes\ket{0},\,\ket{0}\otimes\ket{1}$ and
$\ket{0}\otimes\ket{0}$. One can check that for density matrices
of the form (12)
\begin{equation}
C(\ro)=|\ro_{23}|-\sqrt{\ro_{22}\ro_{33}}-|\,|\ro_{23}|-\sqrt{\ro_{22}\ro_{33}}\,|
\end{equation}
By positive-definiteness of $\ro$, $|\ro_{23}|\leq
\sqrt{\ro_{22}\ro_{33}}$, and we have the result:\\[2mm]
\textit{Concurrence of a density matrix} (12) \textit{is given by}
\begin{equation}
C(\ro)=2\,|\ro_{23}|
\end{equation}
As we will show, the class of density matrices given by (12) is
invariant with respect to the time evolution considered in the
paper, and formula (14) can be used to analyse the evolution of
entanglement of states which  initially have the form (12).
\subsection{Violation of Bell inequalities}
The contradiction between quantum theory and local realism
expressed by the violation of Bell - CHSH inequality \cite{CHSH},
can be studied in the case of two - qubit system using simple
necessary and sufficient condition \cite{HHH,H}. Any state
$\ro\in\fE_{AB}$ can be written as
\begin{equation}
\ro=\frac{1}{4}\left(\I_{2}\otimes\I_{2}+\tl{r}\cdot\tl{\sigma}\otimes\I_{2}+\I_{2}\otimes
\tl{s}\cdot
\tl{\sigma}+\sum\limits_{n,m=1}^{3}t_{nm}\,\si{n}\otimes\si{m}\right)
\end{equation}
where $\I_{2}$ is the identity matrix in two dimensions,
$\si{1},\si{2},\si{3}$ are Pauli matrices, $\tl{r},\tl{s}$ are
vectors in $\R^{3}$ and
$\tl{r}\cdot\tl{\sigma}=\sum\limits_{j=1}^{3}r_{j}\si{j}$. The
coefficients
\begin{equation}
t_{nm}=\tr (\ro\,\si{n}\otimes\si{m})
\end{equation}
form a real matrix $T_{\ro}$. Define also real symmetric matrix
\begin{equation}
U_{\ro}=T_{\ro}^{T}\,T_{\ro}
\end{equation}
where $T_{\ro}^{T}$ is the transposition of $T_{\ro}$.
Consider now the family of
Bell operators
\begin{equation}
B_{CHSH}=\tl{a}\cdot\tl{\sigma}\otimes
(\tl{b}+\tl{b}^{\prime})\cdot\tl{\sigma}+\tl{a}^{\prime}\cdot\tl{\sigma}\otimes
(\tl{b}-\tl{b}^{\prime})\cdot\tl{\sigma}
\end{equation}
where $\tl{a},\,\tl{a}^{\prime},\,\tl{b},\,\tl{b}^{\prime}$ are
unit vectors in $\R^{3}$. Then CHSH inequality reads
\begin{equation}
|\tr (\ro\, B_{CHSH})|\leq 2
\end{equation}
Violation of inequality (19) by the density matrix (15)  and some
Bell operator (18) can be checked by the following criterion: Let
\begin{equation}
m(\ro)=\max_{j<k}\; (u_{j}+u_{k})
\end{equation}
and $u_{j},\, j=1,2,3$ are the eigenvalues of $U_{\ro}$. As was
shown in \cite{HHH,H}
\begin{equation}
\max_{B_{CHSH}}\,\tr (\ro\, B_{CHSH})=2\,\sqrt{m(\ro)}
\end{equation}
Thus (19) is violated by some choice of
$\tl{a},\tl{a}^{\prime},\tl{b},\tl{b}^{\prime}$ iff $m(\ro)>1$.
\par
\noindent
If we consider subclass of the class of states (12)
consisting of density matrices of the form
\begin{equation}
\ro=\begin{pmatrix}0&0&0&0\\
0&\ro_{22}&\ro_{23}&0\\
0&\ro_{32}&\ro_{33}&0\\
0&0&0&\ro_{44}
\end{pmatrix}
\end{equation}
then we obtain the following expression for $m(\ro)$
\begin{equation}
m(\ro)=\max\; (2\,C^{2}(\ro),\,(1-2\,\ro_{44})^{2}+C^{2}(\ro)\,)
\end{equation}
where $C(\ro)=2|\ro_{23}| $ is the concurrence of the state $\ro$.
Notice that the inequality
$$
(1-2\ro_{44})^{2}+C^{2}(\ro)>1
$$
is equivalent to
$$
|\ro_{23}|^{2}>\ro_{44}\,(1-\ro_{44})
$$
Let us introduce  linear entropy of the state $\ro$
$$
S_{L}(\ro)=1-\tr\,\ro^{2}
$$
For states (22)
$$
\tr\,\ro^{2}=\ro_{22}^{2}+\ro_{33}^{2}+\ro_{44}^{2}+2\,|\ro_{23}\,|^{2}
$$
so using $(\ro_{22}+\ro_{33}+\ro_{44})^{2}=1$ we obtain
$$
S_{L}(\ro)=2\,(\ro_{22}\ro_{33}+\ro_{22}\ro_{44}+\ro_{33}\ro_{44}-|\ro_{23}|^{2}\,)
$$
On the other hand
$$
|\ro_{23}|^{2}-\ro_{44}\,(\ro_{22}+\ro_{33}\,)=|\ro_{23}|^{2}-\ro_{44}\,(1-\ro_{44}\,)>0
$$
so
$$
\ro_{22}\ro_{33}-\frac{1}{2}\,S_{L}(\ro)=|\ro_{23}|^{2}-\ro_{44}
\,(\ro_{22}+\ro_{33}\,)>0
$$
and we obtain the following
result: \\[4mm]
\textit{The states} \rm{(22)} \textit{ violate some Bell - CHSH
inequality if and only if $|\ro_{23}|>\frac{1}{2\sqrt{2}}$ or
$\ro_{22}\ro_{33}>\frac{1}{2}\,S_{L}(\ro)$.}
\newpage
\noindent
\section{Spontaneous Emission and Evolution of Entanglement}
We study time evolution of the system of two two-level atoms
separated by a distance $R$ when we take into account only the
dissipative process of spontaneous emission. The dynamics of such
system is given by the master equation \cite{Agar,FT1}
\begin{equation}
\frac{d\ro}{dt}=L_{D}\ro,\quad \ro\in \fE_{AB}
\end{equation}
with the following generator $L_{D}$
\begin{equation}
L_{D}\ro=\frac{1}{2}\sum\limits_{k,l=A,B}\ga_{kl}\,
(\,2\sa{-}{k}\ro\sa{+}{l}-\sa{+}{k}\sa{-}{l}\ro-\ro\sa{+}{k}\sa{-}{l}\,)
\end{equation}
where
\begin{equation}
\sa{\pm}{A}=\si{\pm}\otimes\I,\; \sa{\pm}{
B}=\I\otimes \si{\pm},\; \si{\pm}=\frac{1}{2}(\si{1}\pm i \si{2})
\end{equation}
and $\ga_{AA}=\ga_{BB}=\g,\;\ga_{AB}=\ga_{BA}=\ga=g\g$. Here $\gamma_{0}$
is the single atom spontaneous emission rate,
and $\gamma=g\gamma_{0}$  is a relaxation constant of photon
exchange. In the model, $g$ is the function of the  distance $R$
between atoms and $g\to 1$ when $R\to 0$.  In this section we
investigate the time evolution of the initial density matrix
$\ro$ of the compound system, governed by the semi - group
$\{T_{t}\}_{t\geq 0}$ generated by $L_{D}$. In particular, we will
study the time development of entanglement of $\ro$, measured by
concurrence. When $\ga <\g$, the semi-group $\{T_{t}\}_{t\geq 0}$
is uniquely relaxing, with the asymptotic state
$\ket{0}\otimes\ket{0}$. Thus, for any initial state $\ro$, the
concurrence $C(\ro_{t})$ approaches $0$ when $t\to \infty$. But
still there can be some transient entanglement between atoms
\cite{J,FT3,FT2}. In this section we study in details time evolution
of a given initial state $\ro= (\ro_{jk})$. Direct calculations
show that the state $\ro(t)$ at time $t$ has the following matrix
elements with respect to the basis $\ket{1}\otimes \ket{1},\,
\ket{1}\otimes\ket{0},\, \ket{0}\otimes\ket{1},\,
\ket{0}\otimes\ket{0}$
\begin{equation*}
\begin{split}
\ro_{11}(t)=&e^{-2\g t}\ro_{11}\\[2mm]
\ro_{12}(t)=&e^{-\frac{3}{2}\g t}\,(\ro_{12}\cosh \frac{\ga
t}{2}-\ro_{13}\sinh\frac{\ga t}{2}\,)\\[2mm]
\ro_{13}(t)=&e^{-\frac{3}{2}\g t}\,(\ro_{13}\cosh \frac{\ga
t}{2}-\ro_{12}\sinh\frac{\ga t}{2}\,)\\[2mm]
\ro_{14}(t)=&e^{-\g t}\ro_{14}\\[2mm]
\ro_{22}(t)=&-e^{-2\g
t}\frac{\ga^{2}+\g^{2}}{\g^{2}-\ga^{2}}\,\ro_{11}+ e^{-\g t}\bigg[
\frac{1}{2}(\ro_{22}-\ro_{33})+
\big(\frac{\ga^{2}+\g^{2}}{\g^{2}-\ga^{2}}\,\ro_{11}+\frac{1}{2}\,(\ro_{22}+\ro_{33})\big)\cosh
\ga t \\[2mm]
&-\frac{2\ga
\g}{\g^{2}-\ga^{2}}\,\ro_{11}-\mathrm{Re}\ro_{23}\sinh
\ga t )\bigg ]\\[2mm]
\ro_{33}(t)=&-e^{-2\g
t}\frac{\ga^{2}+\g^{2}}{\g^{2}-\ga^{2}}\,\ro_{11}+ e^{-\g t}\bigg[
\frac{1}{2}(\ro_{33}-\ro_{22})+
\big(\frac{\ga^{2}+\g^{2}}{\g^{2}-\ga^{2}}\,\ro_{11}+\frac{1}{2}\,(\ro_{22}+\ro_{33})\big)\cosh
\ga t \\[2mm]
&-\frac{2\ga
\g}{\g^{2}-\ga^{2}}\,\ro_{11}-\mathrm{Re}\ro_{23}\sinh
\ga t )\bigg ]
\end{split}
\end{equation*}
\begin{equation*}
\begin{split}
\ro_{23}(t)=&-e^{-2\g t}\frac{2\ga
\g}{\g^{2}-\ga^{2}}\,\ro_{11}+e^{-\g t}\,\bigg [
-\big(\frac{\ga^{2}+\g^{2}}{\g^{2}-\ga^{2}}\sinh \ga t+\cosh \ga
t\big)\,\ro_{11}+ i\mathrm{Im}\ro_{23}+\mathrm{Re}\ro_{23}\cosh\ga
t\\[2mm]
& -\frac{1}{2}(\ro_{22}+\ro_{33})\sinh \ga t\bigg]\\[2mm]
\ro_{24}(t)=&\big[e^{-\frac{3}{2}\g t}\,(\sinh \frac{\ga}{2}t-\ga
\cosh\frac{\ga}{2}t)+e^{-\frac{\g}{2}t}\,(\ga
\cosh\frac{\ga}{2}t-\sinh\frac{\ga}{2}t)\big]\ro_{12}+\\[2mm]
&\big[e^{-\frac{3}{2}\g t}\,(\ga\sinh \frac{\ga}{2}t-
\cosh\frac{\ga}{2}t)+e^{-\frac{\g}{2}t}\,(
\cosh\frac{\ga}{2}t-\ga\sinh\frac{\ga}{2}t)\big]\ro_{13}+\\[2mm]
&e^{-\frac{\g}{2}t}\big(\ro_{24}\, \cosh\frac{\ga}{2}t
-\ro_{34}\,\sinh\frac{\ga}{2}t\big)\\[2mm]
\ro_{34}(t)=&\big[e^{-\frac{3}{2}\g t}\,(\ga\sinh \frac{\ga}{2}t-
\cosh\frac{\ga}{2}t)+e^{-\frac{\g}{2}t}\,(
\cosh\frac{\ga}{2}t-\ga\sinh\frac{\ga}{2}t)\big]\ro_{12}+\\[2mm]
&\big[e^{-\frac{3}{2}\g t}\,(\sinh \frac{\ga}{2}t-
\ga\cosh\frac{\ga}{2}t)+e^{-\frac{\g}{2}t}\,(
\ga\cosh\frac{\ga}{2}t-\sinh\frac{\ga}{2}t)\big]\ro_{13}+\\[2mm]
&e^{-\frac{\g}{2}t}\big(\ro_{34}\, \cosh\frac{\ga}{2}t
-\ro_{24}\,\sinh\frac{\ga}{2}t\big)\\[2mm]
\ro_{44}(t)=&1+e^{-2\g
t}\frac{\g^{2}+3\ga^{2}}{\g^{2}-\ga^{2}}\ro_{11}+e^{-\g
t}\,\bigg[\big(\frac{4\ga\g}{\g^{2}-\ga^{2}}\sinh \ga
t-\frac{2(\ga^{2}+\g^{2})}{\g^{2}-\ga^{2}}\cosh \ga t\big
)\ro_{11}\\[2mm]
&-(\ro_{22}+\ro_{33})\cosh\ga t +2\mathrm{Re}\ro_{23}\,\sinh\ga
t\bigg]
\end{split}
\end{equation*}
The remaining matrix elements can be obtained by the Hermiticity
condition. One can simply check that the classes of states (12)
and (22) are invariant with respect to the above time evolution.
In particular, in both cases
\begin{equation}
\ro_{23}(t)=e^{-\g t}\,\big[\,\mathrm{Re}\, \ro_{23}\cosh\ga
t+i\,\mathrm{Im}\,\ro_{23}-\frac{1}{2}\,(\ro_{22}+\ro_{33})\sinh\ga
t\,\big]
\end{equation}
So we obtain the result:\\[2mm]
\textit{For initial states} (12) \textit{or} (22) \textit{the
concurrence at time $t$ is given by
\begin{equation}
C(\ro(t))=2\,e^{-\g t}\,\big|\,\big(\, \mathrm{Re}\,
\ro_{23}\cosh\ga
t+i\,\mathrm{Im}\,\ro_{23}-\frac{1}{2}\,(\ro_{22}+\ro_{33})\sinh\ga
t\,\big)\big|
\end{equation}
where $\ro_{jk}$ are matrix elements of the initial state.}\\[2mm]
Let us consider pure initial states $\Psi\in \C^{4}$ belonging to
the class (12). The most general pure state of this type can be
written as
\begin{equation}
\Psi=\cos\phi\cos\psi\,\ket{1}\otimes\ket{0}+
\sin\phi\cos\psi\,e^{i\Theta}\,\ket{0}\otimes\ket{1}+
\sin\psi\,e^{i\Xi}\,\ket{0}\otimes\ket{0}
\end{equation}
with $\phi,\,\psi\in [0,\frac{\pi}{2}],\; \Theta,\,\Xi\in
[0,2\pi]$. Using (14) we see that concurrence of (29) is given by
\begin{equation}
C(\Psi)=\cos^{2}\psi\,\sin 2\phi
\end{equation}
By (28), the time evolution of this initial concurrence is
described by the following function
\begin{equation}
C(\ro(t))=e^{-\g t}\,\cos^{2}\psi\, \big| \sin 2\phi\, \cos\Theta
\,\cosh \ga t -\sinh\ga t -i \sin2\phi\,\sin \Theta\,\big|
\end{equation}
The function (31) is simple to analyse when
$\phi=0$ or $\frac{\pi}{2}$. In this case
\begin{equation}
\Psi=e^{i\Theta}\,\cos\psi\,\ket{0}\otimes\ket{1}+
e^{i\Xi}\,\sin\psi\,\ket{0}\otimes\ket{0}
\end{equation}
and
\begin{equation}
C(\ro(t))=e^{-\g t}\, \cos^{2}\psi\,\sinh\ga t
\end{equation}
From the formula (33) we see that initial concurrence equal to
zero, increases in the time interval $[0,t_{\mathrm{max}}]$, where
\begin{equation}
t_{\mathrm{max}}=\frac{1}{2\ga}\ln\frac{\g+\ga}{\g-\ga}
\end{equation}
to the maximal value
\begin{equation}
C_{\mathrm{max}}=\cos^{2}\psi\,\frac{\ga}{\g-\ga}
\left(\frac{\g+\ga}{\g-\ga}\right)^{-\frac{\g+\ga}{2\ga}}
\end{equation}
and then asymptotically goes to zero (see \textbf{Fig. 1}).\\[8mm]
\begin{picture}(300,300)
\put(50,70){\begin{picture}(180,180) \epsffile{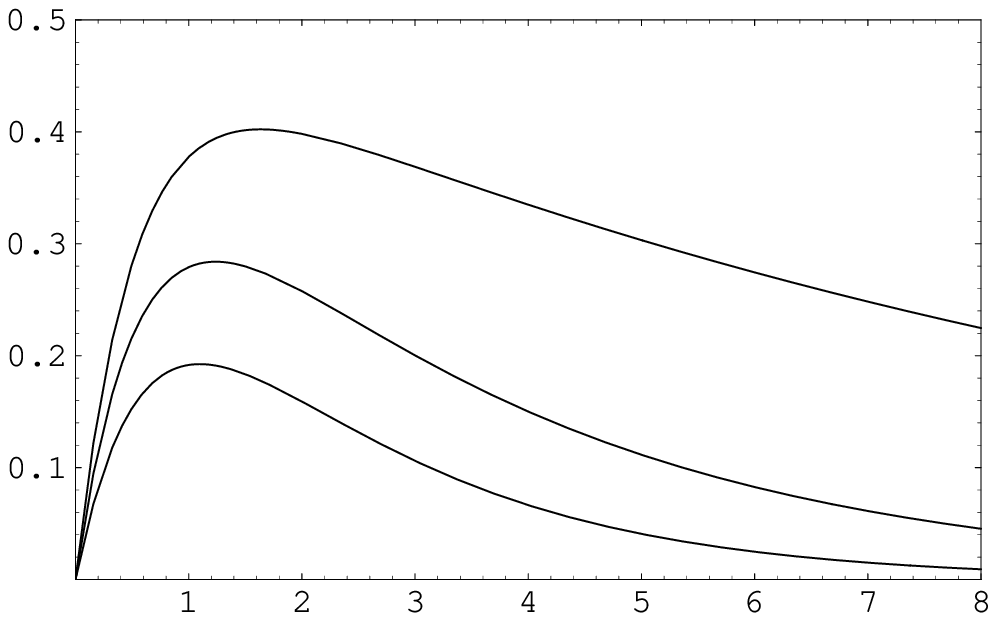}
\end{picture}}
\put(350,75){$\g t$}
\put(60,255){$C(\ro(t))$}
\put(120,150){$0.5$}
\put(130,180){$0.7$}
\put(140,220){$0.9$}
\end{picture}
\vskip -18mm
\noindent \centerline{\textbf{Fig. 1.} $C(\ro(t))$ for initial
states (32) with $\psi=0$ and different values of $\ga/\g$}
\vskip 8mm
\noindent
If $\phi$ is arbitrary, we can put for simplicity $\psi=0$. Then
\begin{equation}
\Psi=\cos\phi
\,\ket{1}\otimes\ket{0}+\sin\phi\,e^{i\Theta}\,\ket{0}\otimes\ket{1}
\end{equation}
and $C(\Psi)=\sin 2\phi$. The evolution of this initial
concurrence is given by
\begin{equation}
C(\ro(t))=e^{-\g t}\,\big |\,\sin2\phi\,\cos\Theta\,\cosh\ga t
-\sinh\ga t -i\,\sin2\phi\,\sin\Theta\,\big|
\end{equation}
Depending on values of $\phi$ and $\Theta$, the function (37) can
be strictly decreasing to zero, or can have one maximal value for some
$t$, or even can have one minimal and one maximal value. Let us
discuss all these possibilities by choosing some special initial
states from the class (36).\\[2mm]
\textbf{a.} Let $\Theta=0$. Then
\begin{equation}
\Psi=\cos\phi\,
\ket{1}\otimes\ket{0}+\sin\phi\,\ket{0}\otimes\ket{1}
\end{equation}
and
\begin{equation}
C(\ro(t))=e^{-\g t}\,\big|\, \sin 2\phi\, \cosh\ga t -\sinh \ga
t\,\big|
\end{equation}
The function (39) is decreasing to zero in the interval
$[0,t_{\mr{min}}]$ where
\begin{equation}
t_{\mr{min}}=\frac{1}{2\ga}\,\ln\frac{1+\sin 2\phi}{1-\sin 2\phi}
\end{equation}
Then in the interval $[t_{\mr{min}},t_{\mr{max}}]$, with
\begin{equation}
t_{\mr{max}}=\frac{1}{2\ga}\,\ln\frac{(1+\sin
2\phi)(\g+\ga)}{(1-\sin 2\phi)(\g-\ga)}
\end{equation}
(39) increases to the maximal value
\begin{equation}
C_{\mr{max}}=\frac{\ga\,|\cos
2\phi|}{\sqrt{\g^{2}-\ga^{2}}}\,\left(\frac{(1+\sin
2\phi)(\g+\ga)}{(1-\sin 2\phi)(\g-\ga)}\right)^{-\frac{\g}{2\ga}}
\end{equation}
For $t>t_{\mr{max}}$, (39) goes asymptotically to $0$. For
sufficiently small initial concurrence $C_{\mr{max}}>C(\Psi)$ but
for larger entanglement of the initial state, the maximal
entanglement produced
during the evolution is smaller then $C(\Psi)$ (see \textbf{Fig. 2.} below). \\[8mm]
\begin{picture}(300,300)
\put(50,70){\begin{picture}(180,180) \epsffile{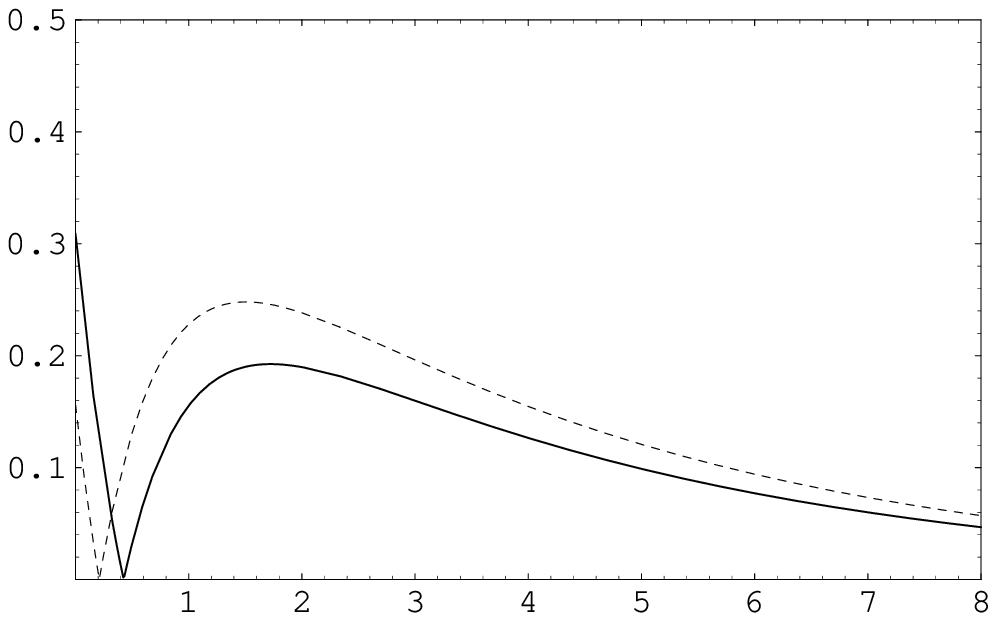}
\end{picture}}
\put(350,75){$\g t$}
\put(60,255){$C(\ro(t))$}
\put(120,120){$\phi=\pi/20$}
\put(140,165){$\phi=\pi/40$}
\end{picture}
\vskip -18mm \noindent \centerline{\textbf{Fig. 2.} $C(\ro(t))$
for initial states (38), with $\phi=\pi/40,\,\pi/20$ and
$\ga/\g=0.75$} \vskip 8mm \noindent \textbf{b.} Let $\Theta=\pi$.
Then
\begin{equation}
\Psi=\cos\phi\,\ket{1}\otimes\ket{0}-\sin\phi\,\ket{0}\otimes\ket{1}
\end{equation}
and
\begin{equation}
C(\ro(t))=e^{-\g t}\, \big|\, \sin 2\phi\,\cosh\ga t+\sinh\ga
t\,\big|
\end{equation}
If $\sin 2\phi\geq \frac{\ga}{\g}$ then function (44) is
monotonically decreasing to $0$. On the other hand, if $\sin 2\phi
<\frac{\ga}{\g}$ then at time
\begin{equation}
t_{\mr{max}}=\frac{1}{2\ga}\,\ln\frac{(1-\sin
2\phi)(\g+\ga)}{(1+\sin 2\phi)(\g-\ga)}
\end{equation}
(44) attains local maximum
\begin{equation}
C_{\mr{max}}=\frac{\ga\,|\cos
2\phi|}{\sqrt{\g^{2}-\ga^{2}}}\,\left(\frac{(1-\sin
2\phi)(\g+\ga)}{(1+\sin 2\phi)(\g-\ga)}\right)^{-\frac{\g}{2\ga}}
\end{equation}
$C_{\mr{max}}$ is always greater then initial concurrence
$C(\Psi)$ and
\begin{equation}
C_{\mr{max}}\to\frac{1+\sin 2\phi}{2}\quad\text{when}\quad
\ga\to\g
\end{equation}
and
\begin{equation}
C_{\mr{max}}\to \frac{\ga}{\g}\quad{when}\quad \sin 2\phi\to
\frac{\ga}{\g}
\end{equation}
Thus, for entanged pure initial states  (43) the dissipative
process of spontaneous emission increases entanglement, provided
that the initial entanglement was smaller then
$\frac{\ga}{\g}$ (see \textbf{Fig. 3}).\\[8mm]
\noindent
\begin{picture}(300,300)
\put(50,70){\begin{picture}(180,180) \epsffile{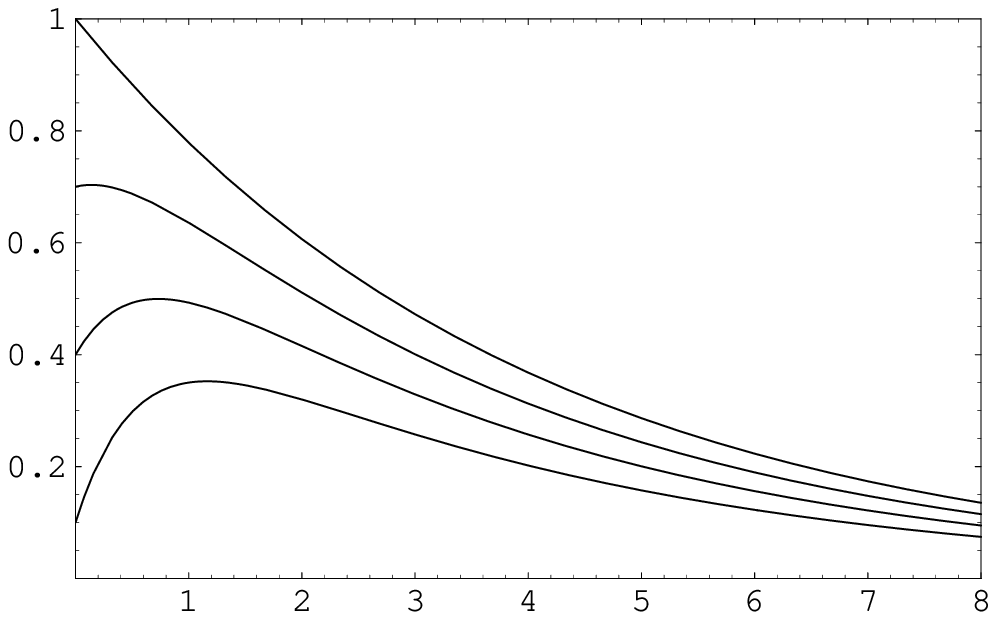}
\end{picture}}
\put(350,75){$\g t$}
\put(60,255){$ C(\ro(t))$}
\put(85,112){$0.1$}
\put(82,145){$0.4$}
\put(79,180){$0.7$}
\put(85,210){$1$}
\end{picture}
\vskip -18mm \noindent \centerline{\textbf{Fig. 3.} $C(\ro(t))$
for initial states (43) with $C(\Psi)=0.1,\,0.4,\,0.7,\,1$ and
$\ga/\g=0.75$}
\vskip 8mm \noindent
\textbf{c.} Let
$\Theta=\pi/2$. Then
\begin{equation}
\Psi=\cos\phi\,\ket{1}\otimes\ket{0}+i\,\sin\phi\,\ket{0}\otimes\ket{1}
\end{equation}
and
\begin{equation}
C(\ro(t))=e^{-\g t}\, \big|\, \sinh\ga t +i\,\sin 2\phi\,\big|
\end{equation}
One can show that if $|\sin 4\phi|<\ga/\g$ then (50) achieves
local minimum at
\begin{equation}
t_{\mr{min}}=\frac{1}{2\ga}\,\ln\frac{\g\cos 4\phi
-\sqrt{\ga^{2}-\g^{2}\sin^{2} 4\phi}}{\g-\ga}
\end{equation}
and local maximum at
\begin{equation}
t_{\mr{max}}=\frac{1}{2\ga}\,\ln\frac{\g\cos 4\phi
+\sqrt{\ga^{2}-\g^{2}\sin^{2} 4\phi}}{\g-\ga}
\end{equation}
with the corresponding values of concurrence
\begin{equation}
C_{\mr{min}}=\left(\frac{\g\cos 4\phi
-\sqrt{\ga^{2}-\g^{2}\sin^{2}
4\phi}}{\g-\ga}\right)^{-\frac{\g}{2\ga}}\,\sqrt{\frac{\ga^{2}\cos
4\phi -\ga\,\sqrt{\ga^{2}-\g^{2}\sin^{2}
4\phi}}{2(\g^{2}-\ga^{2})}}
\end{equation}
and
\begin{equation}
C_{\mr{max}}=\left(\frac{\g\cos 4\phi
+\sqrt{\ga^{2}-\g^{2}\sin^{2}
4\phi}}{\g-\ga}\right)^{-\frac{\g}{2\ga}}\,\sqrt{\frac{\ga^{2}\cos
4\phi +\ga\,\sqrt{\ga^{2}-\g^{2}\sin^{2}
4\phi}}{2(\g^{2}-\ga^{2})}}
\end{equation}
For other cases, the function (50) is monotonically decreasing to
$0$ (see \textbf{Fig. 4.} below). \vskip 8mm \noindent
\begin{picture}(300,300)
\put(50,70){\begin{picture}(180,180) \epsffile{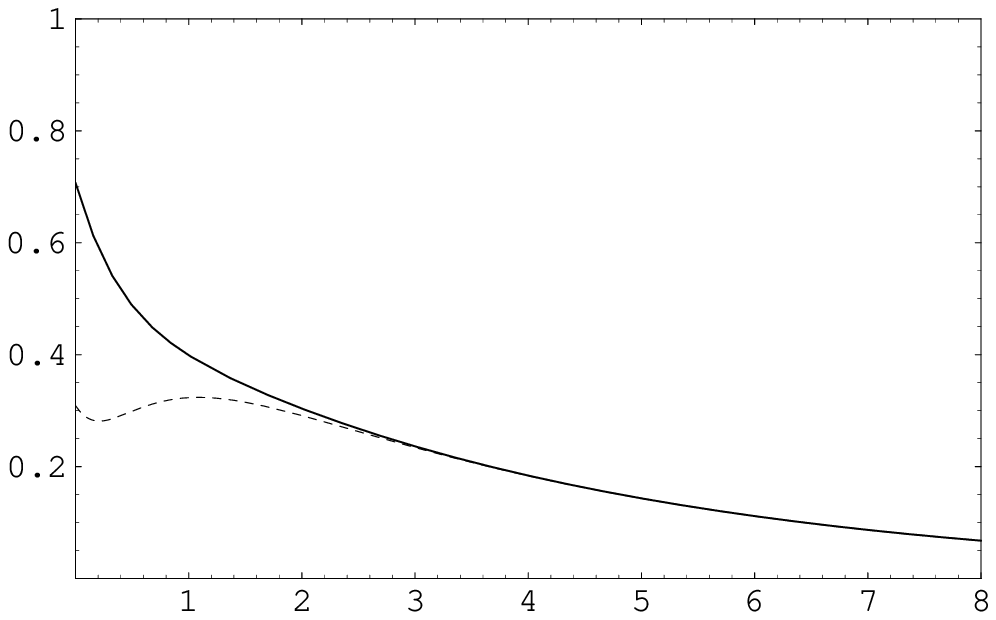}
\end{picture}}
\put(350,75){$\g t$}
\put(60,255){$ C(\ro(t))$}
\put(90,110){$\phi=\pi/20$}
\put(110,160){$\phi=\pi/8$}
\end{picture}
\vskip -18mm \noindent \centerline{\textbf{Fig. 4.} $C(\ro(t))$
for initial states (49) with $\phi=\pi/20,\,\pi/40$ and
$\ga/\g=0.75$} \vskip 8mm \noindent
\newpage
\noindent
\section{Evolution of Nonlocality}
Nonlocality of quantum theory manifesting by violation of Bell
inequalities is strictly connected with the existence of entangled
states. It is known that every pure entangled  state violates some
Bell inequality (see e.g. \cite{Gisin}). But for mixed entangled
states it is no longer true \cite{Werner}. So it is interesting to
discuss how dissipative  process of spontaneous emission
influences  nonlocal properties of initial states. For simplicity
we restrict the class of states considered below to density
matrices of the form (22). For that class we can apply the
results of Sect. 2.2.
\par
\noindent As the initial states we take the states (43). Note
that at time $t$, the state $\ro(t)$ will have the form (22). The
initial entanglement is non-zero, so violation of some Bell -
CHSH inequality occurs at time $t=0$. What happens during the
evolution?  Consider the inequality
\begin{equation}
\ro_{22}(t)\ro_{33}(t)>\frac{1}{2}S_{L}(\ro(t))
\end{equation}
which is sufficient to nonlocality of the state $\ro(t)$. Observe
that $S_{L}(\ro(0))=0$ and $S_{L}(\ro(t))$ is increasing in some
time - interval. On the other hand,
$$
\ro_{22}(0)\ro_{33}(0)\geq
|\ro_{23}(0)|^{2}>0
$$
and $\ro_{22}(t)\ro_{33}(t)$ asymptotically goes to $0$, so there
is some non - empty interval $0\leq t <t_{1}$ for which the
inequality (55) is satisfied. Thus for $0\leq t <t_{1}$, all
states $\ro(t)$ will still have nonlocal properties. We may also
introduce the time $t_{\mr{n}}$ after which nonlocality is lost.
To this end, besides $t_{1}$ consider $t_{2}$ such that
$$
|\ro_{23}(t_{2})|=\frac{1}{2\sqrt{2}}
$$
Then
\begin{equation}
t_{\mr{n}}=\max\, (t_{1},t_{2})
\end{equation}
By the results of Sect. 2.2, all states $\ro(t)$ for $t\geq
t_{\mr{n}}$ will admit local hidden variable model. To illustrate
this concept, consider as initial states   $\Psi^{+}$ and
$\Psi^{-}$ (symmetric and antsymmetric states)
$$
\Psi^{\pm}=\frac{1}{\sqrt{2}}\,\left(
\ket{1}\otimes\ket{0}\pm\ket{0}\otimes\ket{1}\right)
$$
Let $\ro^{\pm}(t)$ denote corresponding density matrices at time $t$. Then
$$
|\ro_{23}^{\pm}(t)|=\frac{1}{2}\,e^{-(\g \pm\ga)t}
$$
and
$$
\ro_{22}^{\pm}(t)\ro_{33}^{\pm}(t)=\frac{1}{4}\,e^{-2(\g\pm\ga)t},\quad
S_{L}(\ro^{\pm}(t))=2\left(e^{-(\g\pm\ga)t}-e^{-2(\g\pm\ga)t}\right)
$$
We see that
$$
t_{1}=\frac{1}{\g\pm \ga}\;\ln\frac{5}{4},\quad
t_{2}=\frac{1}{\g\pm \ga}\;\frac{\ln 2}{2}
$$
Thus
$$
t_{\mr{n}}=\frac{1}{\g\pm \ga}\;\frac{\ln 2}{2}
$$
Note that for antisymmetric state $\Psi^{-}$ and $\ga$ close to
$\g$, $t_{\mr{n}}$ goes to infinity.
\par
\noindent It is also interesting to study in more details time
evolution of the measure of nonlocality which may be defined as follows
\begin{equation}
n(\ro)=\max\, (\,0,\,m(\ro)-1\,)
\end{equation}
As is
well known, $m(\ro)\leq 2$, so $0\leq n(\ro)\leq 1$ and larger value of
$n(\ro)$ gretar then $1$ means violation of CHSH inequality to a larger extent.
Since $m(\ro)=2$ for maximally entangled pure states which maximally
violate CHSH inequalities, for them $n(\ro)=1$. To obtain analytic expression for
$n(\ro(t))$ we can utilize formula (23). Formula (23) is further
simplified if we take such initial states that
\begin{equation}
|\ro_{23}(t)|<\frac{1}{2\sqrt{2}}
\end{equation}
for all $t$. This condition can be achieved if
$$
\frac{\ga}{\g}<\frac{1}{\sqrt{2}}
$$
Then  inequality (58) is satisfied and $m(\ro(t))>1$ if and only
if
\begin{equation}
(1-2\ro_{44}(t))^{2}+C^{2}(\ro(t))>1
\end{equation}
and to study the time evolution of nonlocality, we only need to
know time-dependence of the left hand side of (59). If we take
initial state with $\sin 2\phi<\frac{\ga}{\g}<\frac{1}{\sqrt{2}}$,
then during the time evolution $C(\ro(t))$  increases to the
maximal value $C_{\mr{max}}<\frac{1}{\sqrt{2}}$ in the  interval
$[0,t_{\mr{max}}]$. At the same time,
\begin{equation}
(1-2\ro_{44})^{2}=\left(\,2e^{-\g t}\,( \cosh\ga t+\sin
2\phi\,\sinh\ga t) -1\,\right)^{2}
\end{equation}
decreases so fast that left hand side of (59) is a decreasing
function of $t$ (see \textbf{Fig. 5, 6} below).
Thus we obtain the result:\\[2mm]
\textit{For initial states} (43) \textit{ and $\sin
2\phi<\frac{\ga}{\g}<\frac{1}{\sqrt{2}}$,
nonlocality of $\ro(t)$ given by $n(\ro(t))$ decreases during the time
evolution even if the entanglement increases.}
\\[4mm]
\begin{picture}(300,300)
\put(50,70){\begin{picture}(180,180) \epsffile{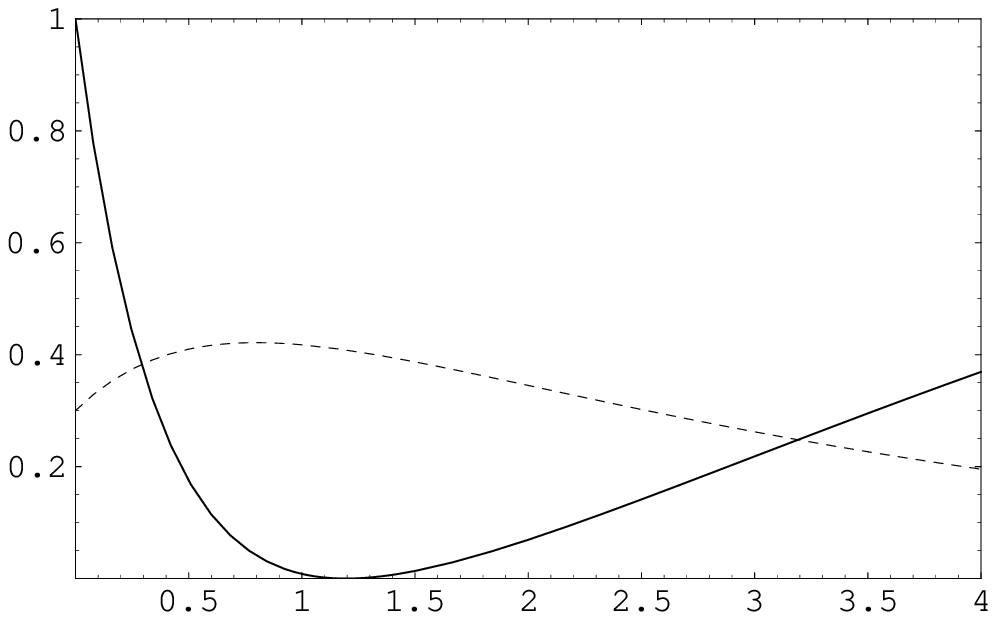}
\end{picture}}
\put(350,75){$\g t$}
\end{picture}
\vskip -18mm \noindent \centerline{\textbf{Fig. 5.} $C(\ro(t))$
(dotted line) and $(1-2\ro_{44}(t))^{2}$ (solid line) for initial
state (43) with $C(\Psi)=0.3$ and $\ga/\g=0.7$}\vskip 4mm
\noindent
\begin{picture}(300,300)
\put(50,70){\begin{picture}(180,180) \epsffile{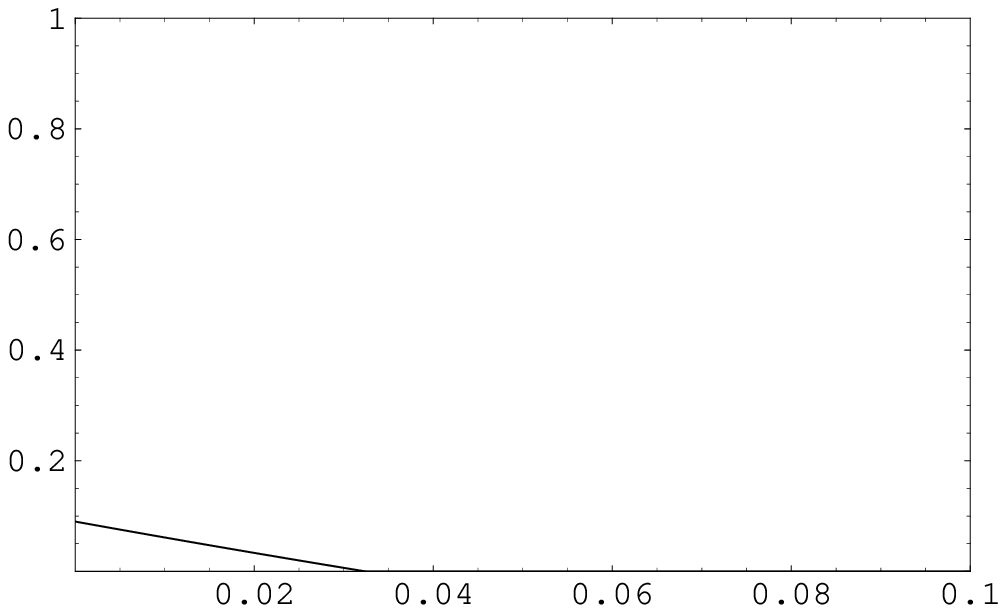}
\end{picture}}
\put(350,75){$\g t$}
\put(60,255){$ n(\ro(t))$}
\end{picture}
\vskip -18mm \noindent \centerline{\textbf{Fig. 6.} $n(\ro(t))$
for initial state (43) with $C(\Psi)=0.3$ and $\ga/\g=0.7$}
\vskip 4mm \noindent
Numerical analysis indicates that for other initial states (43)
for which $m(\ro(t))$ is defined by the whole expression (23), the
nonlocality $n(\ro(t))$ also monotonically goes to $0$
irrespective of the evolution of entanglement (see \textbf{Fig. 7} below).
\vskip 4mm \noindent
\begin{picture}(300,300)
\put(50,70){\begin{picture}(180,180) \epsffile{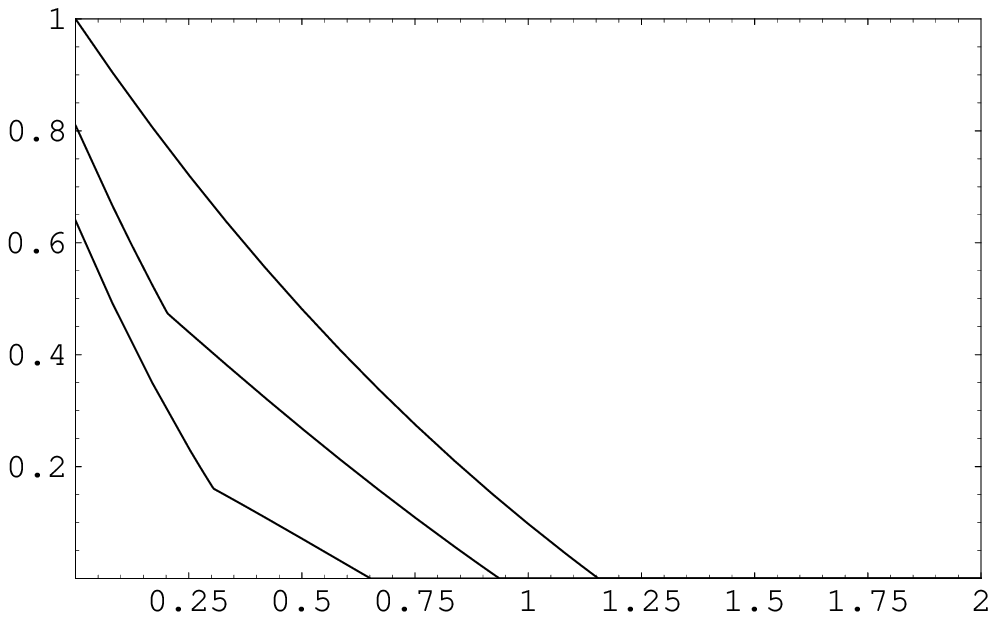}
\end{picture}}
\put(350,75){$\g t$} \put(60,255){$ n(\ro(t))$}
\put(110,130){$0.9$} \put(90,110){$0.8$} \put(130,145){$1.0$}
\end{picture}
\vskip -18mm \noindent \centerline{\textbf{Fig. 7.} $n(\ro(t))$
for initial states (43) with $C(\Psi)=0.8,\,0.9,\,1.0$ and
$\ga/\g=0.7$}

\end{document}